\documentclass{revtex4}
\usepackage[ansinew]{inputenc}
\usepackage[english]{babel}
\usepackage{amsmath,amssymb}
\usepackage{graphicx}

\begin{document}
\title{Numerical solutions of the Dicke Hamiltonian}
\author{Miguel A. Bastarrachea-Magnani, Jorge G. Hirsch}
\affiliation{Instituto de Ciencias Nucleares, Universidad Nacional Aut\'onoma de M\'exico \\ Apdo. Postal 70-543, Mexico D. F., C.P. 04510}

\begin{abstract}
We study the numerical solutions of the Dicke Hamiltonian, which describes a system of many two level atoms interacting with a monochromatic radiation field into a cavity. The Dicke model is an example of a quantum collective behavior which shows superradiant quantum phase transitions in the thermodynamic limit. Results obtained employing two different bases are compared. Both of them use the pseudospin basis to describe the atomic states. For the photon states we use in one case Fock states, while in the other case we use 
a basis built over a particular coherent state, associated to each atomic state.
It is shown that, when the number of atoms increases, the description of the ground state of the system in the superradiant phase requires an equivalent number of photons to be included. This imposes a strong limit to the states that can be calculated using Fock states, while the dimensionality needed to obtain convergent results in the other basis decreases when the atomic number increases, allowing calculations that are very difficult in the Fock basis. Naturally, it reduces also the computing time, economizing computing resources. We show results for the energy, the photon number and the number of excited atoms, for the ground and the first excited state. 
\end{abstract}
\maketitle


\section{Introduction}
The Dicke model describes a system of $N$ two-level atoms interacting with a single monochromatic electromagnetic radiation mode. One of this model's interesting properties is its quantum phase transition in the thermodynamic limit: when the coupling constant between the atoms and the radiations increases, the system undergoes a phase transition between the normal and the superradiant phases. This transition is an example of quantum collective behavior. In recent years the Dicke model has attracted renewed interest in the study of the phase transitions, entanglement in many body systems, chaos in quantum mechanics, and applications in quantum optics and condensed matter physics. At the light of experimental realizations of the superradiant quantum phase transitions \cite{Bau10}, exploring numerical solutions to the model becomes of practical interest. The purpose of this work is to present numerical solutions which make easy solving the system, given that, for finite $N$, only in the limits in which the interaction strength between the atoms and the electromagnetic field becomes null or infinit the system is exactly solvable. 
In particular, we employ coherent states \cite{Che08} to improve the efficiency in obtaining numerical solutions of the Hamiltonian. Expectation values of observables of interest are also presented.

\subsection{Description of atomic-field interaction}
The interaction between $N$ two-level atoms and a multimode radiation field can be described in its more general form by a Hamiltonian formed by three terms: one associated to the radiation modes, one related to the atomic part and a third one which describes the interaction between them\cite{Scu97}.  
\begin{equation}
H=\sum_{k} \hbar \omega_k a^{\dagger}_k a_k + \Delta J_z +\hbar \sum_{k} \frac{\gamma_k}{\sqrt{N}} (a^{\dagger}_k + a_k)(J_+ + J_-)
\end{equation}
Each radiation mode has a frequency $\omega_k$ and it is weighted by the number operator $a^{\dagger}_k a_k$ of this mode. For the atomic part $\Delta$ is the transition frequency, meanwhile $J_z$, $J_+$, $J_-$ are pseudospin collective atomic operators between the ground state $|b \rangle$ and the excited state $|a\rangle$, which obey the SU(2) algebra; they describe the two-level atoms ensemble with an eigenvalue $j$ of length $N/2$. The $J_z = |a\rangle\langle a|-|b\rangle\langle b|$ operator gives the number of atoms in the excited state. The $J_+ = |a\rangle\langle b|$ operator excites the atom and the $J_- = |b\rangle\langle a|$ operator lowers it to the ground state. Finally, the interaction parameter $\gamma_k$ depends principally on the atomic dipolar moment. The description in this work considers a single mode of radiation and it is called the Dicke model. Therefore, the Dicke Hamiltonian can be written as \cite{Dic54} :
\begin{equation}
H_{Dicke}=\omega a^{\dagger} a + \Delta J_z +  \frac{\gamma}{\sqrt{N}} \left(a^{\dagger} + a\right) \left(J_+ + J_-\right)
\end{equation}
There are some well-known approximations of this model. In the so called rotating-wave approximation the terms $a_kJ_-$ and $a^{\dagger}_kJ_+$, which represent interactions beyond the exchange of energy between non-perturbed atoms and the field, are neglected. In this approximation it is called the Tavis-Cummings model \cite{Tav68}. If we consider the simplest case we have only one atom interacting with a single radiation mode. This is called the Jaynes-Cummings model \cite{Jay63}. In this case the $J_z$, $J_+$, $J_-$ operators are commonly written as the $\sigma_z$, $\sigma_+$, $\sigma_-$ operators, which follow the same algebra, but describes only a single two-level atom.

Returning to the Dicke Hamiltonian, the interaction parameter is scaled by $\sqrt{N}$ factor to have the proper thermodynamical limit. 
The Dicke model is exactly solvable in two limits, discussed in the next section. Analytical solutions which are exact in the thermodynamic limit have been found both for the normal \cite{Ema03} and the superradiant \cite{Cas11} phases. In general, for finite $N$ there are no exact solutions. For this reason it is worth to explore numerical solutions. As it was mentioned above, the interest in solving the Dicke model for a finite $N$ comes from the fact that many systems could be described using this Hamiltonian and also it could help to understand more about the phase transitions in quantum systems.

\section{Numerical solutions to the finite Dicke model}
\subsection{The limit $\gamma \rightarrow 0$: Solutions using Fock states}
At zero interactions, the Dicke Hamiltonian eigenstates are the tensor product between Fock states for the radiation modes and angular momentum eigenstates for the atomic part $|\Psi\rangle = |n\rangle |j,m\rangle$, with $j=N/2$. This is possible because the atomic operators are pseudospin ones, as it was mentioned above. The energy spectra in this case is
\begin{equation}
E_{\gamma \rightarrow 0}=\omega n +\frac{1}{2}\Delta m .
\end{equation}

For finite values of $\gamma$ the interaction term mixes states with different number of photons $n$ and different occupation numbers in the two level atoms $j+m$.To diagonalize the Hamiltonian by numerical means it is necessary to truncate the number of photons in the field. 
The Hamiltonian matrix elements are very simple:
\begin{center}
\begin{eqnarray}
%
\langle n^{'};j,m^{'}|H|n;j,m\rangle = 
( n \omega + m \Delta) \delta_{n^{'},n} \delta_{m^{'},m} \nonumber \\
+ \frac{\gamma}{\sqrt{N}} \left(\sqrt{n+1}\,\,\delta_{n^{'},n+1} + \sqrt{n}\,\,\delta_{n^{'},n-1}\right) \\
\times \left(\sqrt{j(j+1)-m(m+1)} \,\,\delta_{m^{'},m+1} + \sqrt{j(j+1)-m(m-1)}\,\,\delta_{m^{'},m-1}\right) \nonumber
\end{eqnarray}
\end{center}

For each eigenstate there exists a number of photons which is enough to ensure the convergence of the solution. So, this value will be the one in which the energy will converge to a correct value and we call this number for the ground state $n_{max}$. The dimension of the Hamiltonian matrix to be diagonalized is $dimH_{Fock}=((n_{max}+1)(2j+1))^{2}$. As $j$ is increased, the number of photons necessary to obtain convergence increases too, turning impracticable the numerical evaluation of the eigenstates. 

The Dicke Hamiltonian has a dynamical symmetry associated with the projectors of the symmetric and antisymmetric representations of the cyclic group $C_2$, given by \cite{Cas11}, 
	\[
	P_S= \frac{1}{2} \bigl( 1 + e^{i \pi \hat{\Lambda}} \bigr) \, , 
	\qquad P_A= \frac{1}{2} \bigl( 1 - e^{i \pi \hat{\Lambda}} \bigr) \, .
	\]
This symmetry allows the classification of the eigenstates in terms of the parity of the eigenvalues $\lambda = j +m + n$ of the excitation number operator 
	\[
	\hat\Lambda = \sqrt{\hat{J}^2 + 1/4} -1/2 + \hat{J}_z + a^{\dagger}a . 
	\]
Under the unitary transformation $U = e^{i \pi \hat{\Lambda}}$; it is immediate that \cite{Cas11} 
	\begin{equation}
		U \, a \,  U^\dagger =  a \,  e^{-i \pi} \, = - a, \qquad  U \, J_{+} \, U^\dagger =  J_{+} \, e^{i \pi}\ = -J_{+}, \qquad U \, J_{z} \, U^\dagger =  J_{z}
		\label{trans}
	\end{equation}
with the corresponding hermitean conjugated expressions. 

It must be remarked that the eigenstates of the Hamiltonian must be simultaneous eigenstates of the parity operators $P_S$ or $P_A$.

\subsection{The limit $\Delta \rightarrow 0$: Solutions using coherent states}

After a rotation of the pseudospin operators by $\frac{\pi}{2}$ around the $y$ axis, the ${J}_z$ operator becomes $-J'_x = -(J'_+ + J'_-)$ and ${J}_x$ becomes $J'_z$. This leaves the Hamiltonian as:
\begin{equation}
H=\omega a^{\dagger} a -\frac{1}{2}\Delta (J'_+ + J'_-) + \frac{2 \, \gamma}{\sqrt{N}} (a^{\dagger} + a) J'_z.
\end{equation}

Then, we build a new annihilation operator, by shifting the original one. This gives us:
\begin{equation}
A = a + \frac{2 \gamma}{\omega \sqrt{N}} J'_z  \equiv a + G J'_z .
\end{equation}  

By substituting $A$ in the Hamiltonian we obtain:
\begin{equation}
H=\omega (A^{\dagger} A - G^{2} {J'_z}^{2}) - \frac{\Delta}{2}(J'_+ + J'_-).
\end{equation}

This Hamiltonian has the same symmetries of the original one.

Using the operators $A, A^\dagger$ a new basis is built, employing coherent states which are functions of the parameter $G$ and the eigenvalues of the $J_z$ operator. We can obtain this states by applying powers of the shifted creation operator acting over the $A$ vacuum state.
\begin{equation}
|\alpha=-Gm\rangle=e^{-{|\alpha|}^{2}/2}\sum^{\infty}_{n=0}\frac{\alpha^{n} A^{n}}{n!}|0\rangle_{b}
\end{equation}    
Now, the states that describe the radiation are a mix of the field and the atomic states, and the interaction is described in terms of $G=\frac{2 \gamma}{\omega \sqrt{N}}$ instead of $\gamma$. 
It would seem that the operators are not coupled, however, this is not true because the operator $A$ contains both $a$ and $J_z$. The matrix's elements in this coherent basis are \cite{Che08}:
\begin{eqnarray}
{}_{b}\langle n^{'};j,m^{'}|H|n;j,m\rangle_{b} = \omega \delta_{n^{'},n} \delta_{m^{'},m}(n-G^{2}m^{2}) \nonumber \\
 - \Delta (\sqrt{j(j+1)-m(m+1)}\,\,\delta_{m^{'},m+1}\, 
{}_{b}\langle n^{'};m^{'}|n;m+1\rangle_{b} \nonumber \\+ \sqrt{j(j+1)-m(m-1)} \,\,\delta_{m^{'},m-1}
\,{}_{b}\langle n^{'};m^{'}|n;m-1\rangle_{b}) 
\end{eqnarray}
The cost we pay for using this basis is that it is not orthogonal anymore, so we have overlaps between states:
\begin{equation}
\langle n^{'};m^{'}|n;m\rangle_{m^{'}>m} =e^{-\frac{G^{2}}{2}} \sum^{min(n^{'},n)}_{k=0} \frac{\sqrt{n^{'}!n!}}{(n^{'}-k)!(n-k)!k!} (-1)^{n^{'}-k} G^{n+n^{'}-2k} ,
\end{equation}
\begin{equation}
\langle n^{'};m^{'}|n;m\rangle_{m^{'}<m} =e^{-\frac{G^{2}}{2}} \sum^{min(n^{'},n)}_{k=0} \frac{\sqrt{n^{'}!n!}}{(n^{'}-k)!(n-k)!k!} (-1)^{n-k} G^{n+n^{'}-2k}.
\end{equation}.

Notice that, if we denote by $n'$ the eigenvalues of $A^{\dagger} A$ and by $m'$ the eigenvalues of $J'_z$, in limit $\Delta \rightarrow 0$ the Hamiltonian is exactly solvable in the new basis, with energies

\begin{equation}
E_{\Delta \rightarrow 0}= \omega (n' - G^{2} {m'}^{2}). 
\end{equation}

In this limit the ground state is doubly degenerate, with $n'=0$ and ${m'}^{2} =j^2$. The eigenstates with the appropriate $C_2$ symmetry are 
$$
|\Delta \rightarrow 0, \pm \rangle = \frac {1} {\sqrt{2}} \left\{|-Gj \rangle |j j\rangle \pm |Gj \rangle |j -j\rangle\right\}
$$.

When $\Delta \neq 0$ for $\gamma >> \omega$ and $\gamma >> \Delta$, the maximum value of $n'$, which we will call $N_{max}$, will be very small. The dimension of the Hamiltonian matrix to be diagonalized in this case is $dimH_{coh}=((N_{max}+1)(2j+1))^{2}$, which can be orders of magnitude smaller than $dimH_{Fock}$, allowing for calculations that in the other basis would have been impossible.

\section{Results}

Numerical solutions are presented for the ground and first excited states in resonance, $\Delta = \omega$, for both basis. We have calculated $n_{max}$ ($N_{max}$), and estimated the computing time in seconds, for several $j$ in resonance employing the ground state; these results are shown in table 1. The energies are measured in  units of $\omega$. The source code was written in Wolfram Mathematica. It allows to find the eigenvalues and eigenvectors of the Hamiltonian, and to evaluate the expectation values of all observables of interest. The computing time is calculated through the Mathematica command ``Timing", with the code running in a computer with a processor AMD Turion 64 X2 at 2.00GHz. While it is clear that the absolute times reported here strongly depend on the platform employed to run the code, the trends and the relative values should be mainly independent of them.

In Fig. 1 we show $n_{max}$ as a function of $\gamma$ for several $j$ for the Fock basis, while in Fig. 2 we plot $N_{max}$ in the coherent state basis. It must be remarked here that the Fock $n_{max}$ is the maximum value that the  $a^{\dagger}a$ eigenvalues can take, while the coherent $N_{max}$ is the same for the $A^{\dagger}A$ eigenvalues, so the Fock $n_{max}$ defines the truncation in the photon number and the coherent $n_{max}$ does not. It is worth to compare them because both are proportional to the dimensionality of the Hamiltonian matrices in each basis. As it is shown in the table and in the figures, for a large number of atoms, the time and computing capacity necessary is too large if we use the Fock basis. In fact, some data are missing for the Fock basis due the long computing times needed to generate them. The comparison between computing time is important in order to economize computing resources, because this permits to calculate systems with more atoms in less time, in other words for large $N$ the coherent basis optimizes the use of the computing resources. It can be seen that the time and number of excitations $N_{max}$ needed for convergence decreases in a significant way if we use instead the coherent basis. 


\begin{table}
\label{Table 1}
\begin{center}
\scalebox{0.8}{\begin{tabular}{|c|c|c|c|c|c|}
\hline
\multicolumn{6}{|c|}{For j=1}\\  \hline
$\gamma$ & Energy & Fock $n_{max}$  & Fock Computing time & Coherent $N_{max}$ & Coherent Computing time \\ \hline
0 & -1.00000 & 2 & 0.015 & 2 & 0.063 \\ \hline
0.1 & -1.00504 & 3 & 0.031 & 3 & 0.081  \\ \hline
0.2 & -1.02062 & 4 & 0.094 & 4 & 0.296\\ \hline
0.5 & -1.15370 & 7 & 0.218 & 8 & 2.200\\ \hline
1.0 & -2.15428 & 13 & 0.967 & 16 & 27.503\\ \hline
2.0 & -8.03226 & 27 & 6.786 & 37 & 806.712\\ \hline
\multicolumn{6}{|c|}{For j=2}\\  \hline
$\gamma$ & Energy & Fock $n_{max}$  & Fock Computing time & Coherent $N_{max}$ & Coherent Computing time \\ \hline
0 & -2.00000 & 2 & 0.011 & 2 & 0.01 \\ \hline
0.1 & -2.00505 & 3 & 0.093 & 3 & 0.265  \\ \hline
0.2 & -2.02085 & 4 & 0.171 & 4 & 0.499\\ \hline
0.5 & -2.17336 & 8 & 0.702 & 8 & 4.430\\ \hline
1.0 & -4.26487 & 19 & 6.692 & 15 & 46.426\\ \hline
2.0 & -16.06350 & 42 & 65.520 & 24 & 296.495\\ \hline
\multicolumn{6}{|c|}{For j=5}\\  \hline
$\gamma$ & Energy & Fock $n_{max}$  & Fock Computing time & Coherent $N_{max}$ & Coherent Computing time \\ \hline
0 & -5.00000 & 2 & 0.218 & 2 & 0.702 \\ \hline
0.1 & -5.00506 & 3 & 0.437 & 3 & 1.888  \\ \hline
0.2 & -5.02099 & 4 & 0.717 & 4 & 3.196\\ \hline
0.5 & -5.19716 & 10 & 7.052 & 8 & 35.007\\ \hline
1.0 & -10.63840 & 30 & 150.354 & 12 & 144.395\\ \hline
2.0 & -40.15720 & 78 & 2499.9600 & 16 & 425.243\\ \hline
\multicolumn{6}{|c|}{For j=10}\\  \hline
$\gamma$ & Energy & Fock $n_{max}$  & Fock Computing time & Coherent $N_{max}$ & Coherent Computing time \\ \hline
0 & -10.00000 & 2 & 0.686 & 2 & 2.137 \\ \hline
0.1 & -10.00510 & 3 & 1.529 & 3 & 5.554  \\ \hline
0.2 & -10.02100 & 5 & 4.446 & 4 & 11.481\\ \hline
0.5 & -10.21300 & 13 & 50.919 & 8 & 102.836\\ \hline
1.0 & -21.26310 & 47 & 2104.800 & 10 & 222.161\\ \hline
2.0 & -80.31340 & -- & -- & 13 & 583.304\\ \hline
\multicolumn{6}{|c|}{For j=20}\\  \hline
$\gamma$ & Energy & Fock $n_{max}$  & Fock Computing time & Coherent $N_{max}$ & Coherent Computing time \\ \hline
0 & -20.00000 & 2 & 2.824 & 2 & 8.206 \\ \hline
0.1 & -20.00510 & 3 & 5.647 & 3 & 22.105  \\ \hline
0.2 & -20.02110 & 4 & 17.831 & 4 & 46.785\\ \hline
0.5 & -20.22690 & 15 & 311.003 & 8 & 424.572\\ \hline
1.0 & -42.51290 & -- & -- & 9 & 635.408\\ \hline
2.0 & -160.62600 & -- & -- & 10 & 888.784\\ \hline
\end{tabular}}
\caption{Computing times and $n_{max}$ ($N_{max}$) for several $j$ as $\gamma$ is increased for both basis.} 
\end{center}
\end{table} 

\begin{center}
\begin{figure}
\includegraphics[scale=0.25]{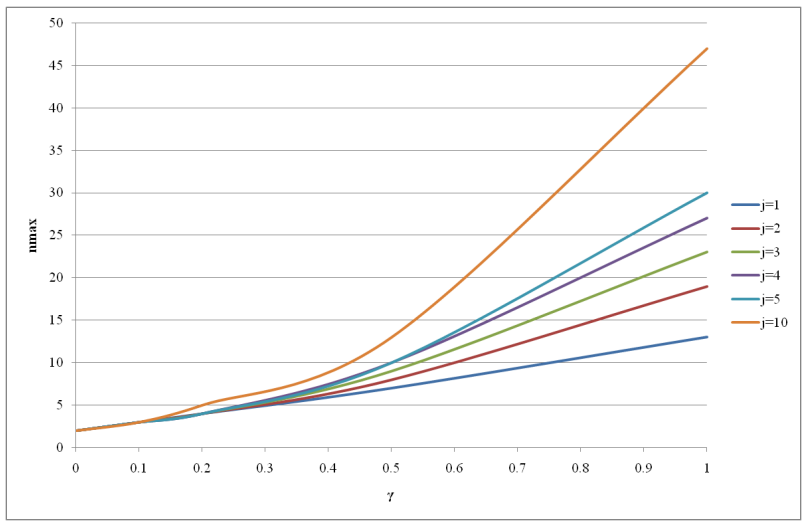}
\caption{Fock $n_{max}$ as function of $\gamma$ for values of $j$ from 1 to 10.}
\label{Figure 1}
\end{figure}
\end{center}

\begin{center}
\begin{figure}
\includegraphics[scale=0.25]{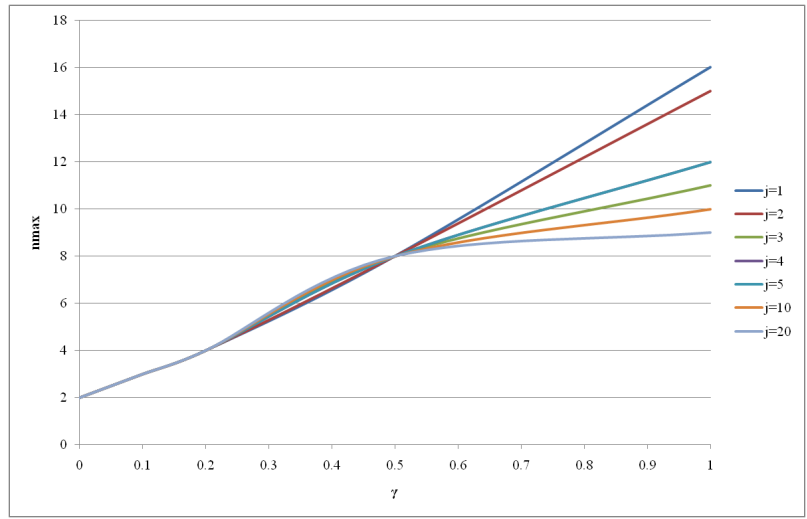}
\caption{Coherent $N_{max}$ as function of $\gamma$ for values of $j$ from 1 to 20.}
\label{Figure 2}
\end{figure}
\end{center}

\begin{center}
\begin{figure}
\includegraphics[scale=1.2]{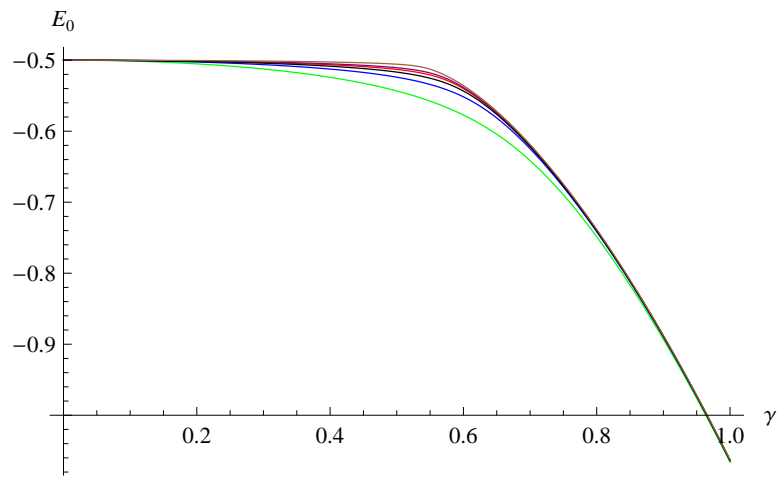}
\caption{The ground state energy as function of $\gamma$, increasing the value of $j$ from 1 to 20.}
\label{Figure 3}
\end{figure}

\begin{figure}
\includegraphics[scale=1.2]{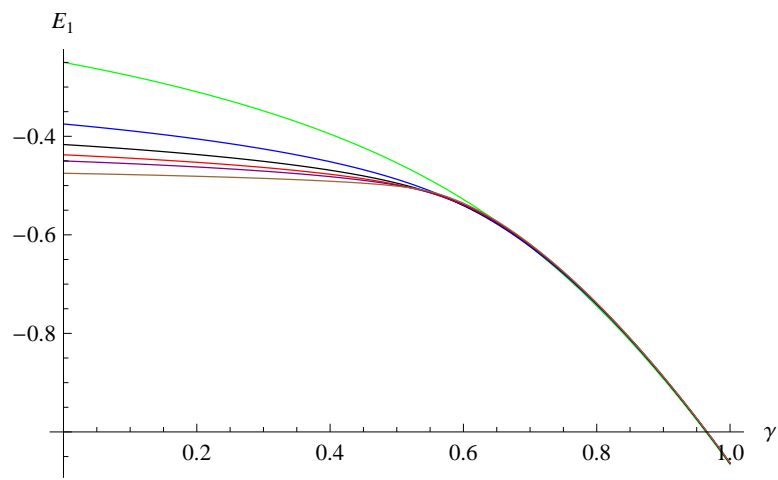}
\caption{First excited state energy as function of $\gamma$, increasing the value of $j$ from 1 to 20.}
\label{Figure 4}
\end{figure}

\begin{figure}
\includegraphics[scale=1.2]{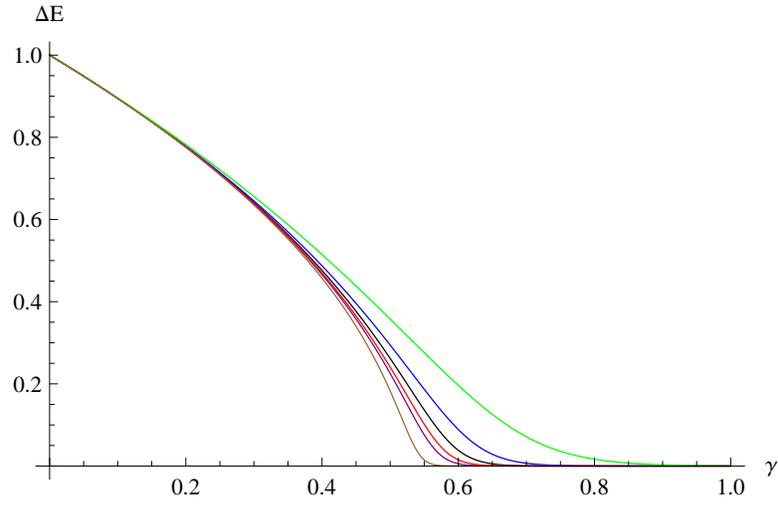}
\caption{Excitation energy as function of $\gamma$, increasing the value of $j$ from 1 to 20.}
\label{Figure 5}
\end{figure}

\begin{figure}
\includegraphics[scale=1.2]{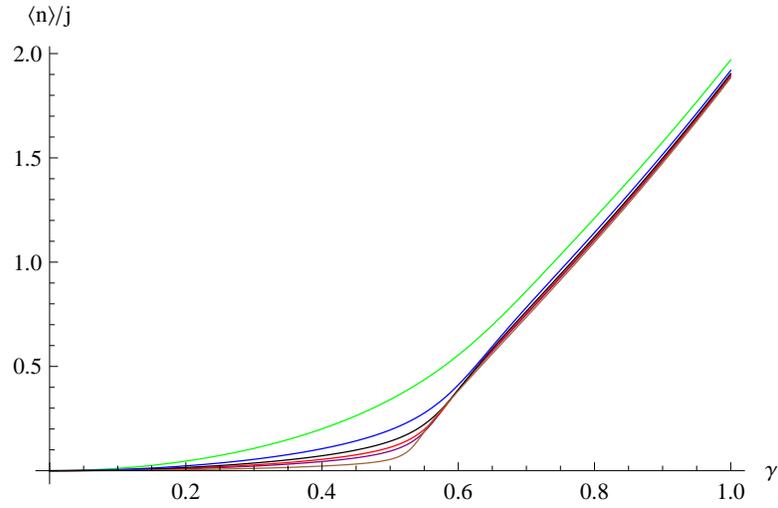}
\caption{Expectation value for the photon number in the ground state as function of $\gamma$, for $j$ from 1 to 20.}
\label{Figure 6}
\end{figure}

\begin{figure}
\includegraphics[scale=1.2]{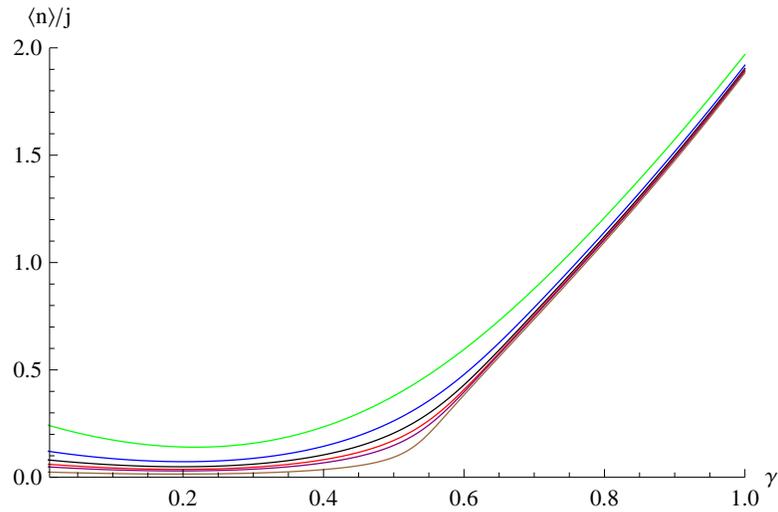}
\caption{Expectation value for the photon number in the first excited state as function of $\gamma$,  for $j$ from 1 to 20.}
\label{Figure 7}
\end{figure}

\begin{figure}
\includegraphics[scale=1.2]{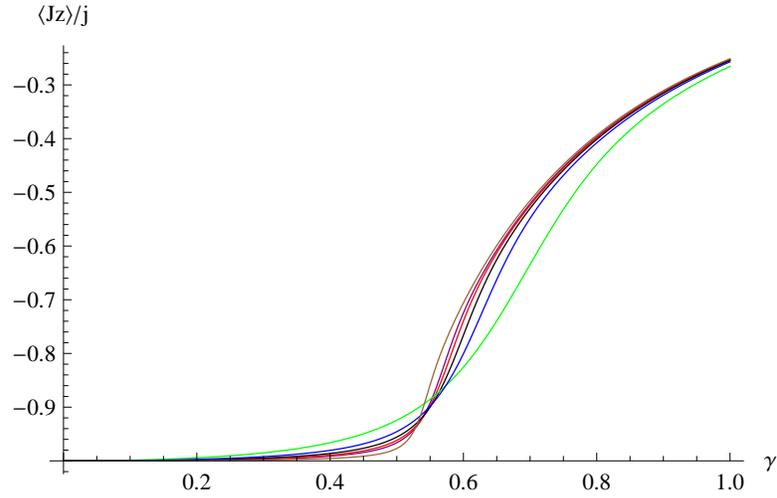}
\caption{Expectation value for $J_z$ for the ground state as function of $\gamma$, for $j$ from 1 to 20.}
\label{Figure 8}
\end{figure}

\begin{figure}
\includegraphics[scale=1.2]{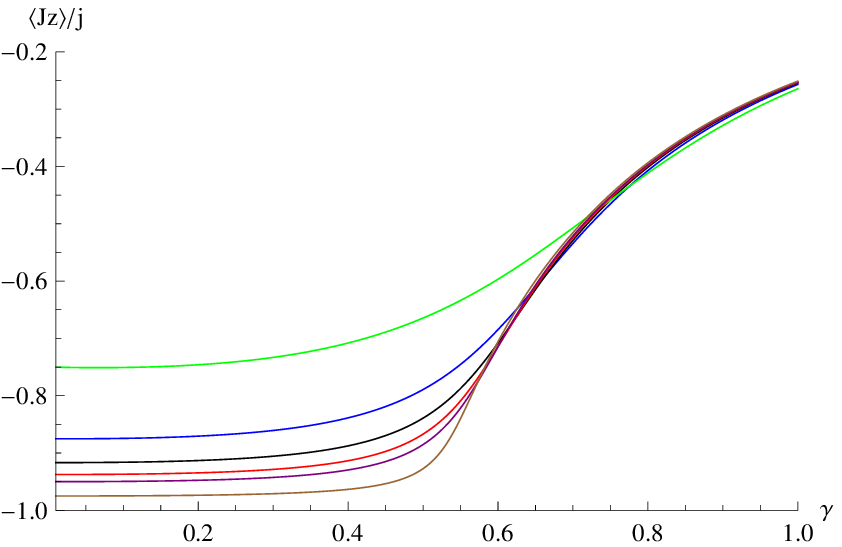}
\caption{Expectation value for $J_z$ for the first excited state as function of $\gamma$, for $j$ from 1 to 20.}
\label{Figure 9}
\end{figure}

\end{center}

In figure 3 and figure 4 we show the ground and first excited energies and in figure 5 the difference between them varying the interaction parameter. Likewise, in figures 6 to 9 it is shown the photon number and number of excitations expected values as functions of $\gamma$. Both observables are divided by $j$ in order to scale them with a system natural size. It can be seen in all the observables that there is a critical value of the interaction parameter where a phase transition takes place. Also, we note the effect is clearer with the increase of $j$, but we do not need large values of $N$ to observe the transition. So, using this basis, we can reproduce the results about the observables shown in \cite{Ema03}. As it was shown in \cite{Cas06}-\cite{2Cas09}, coherent states are a good asymptotic approximation for the states as $j$ tends to infinity and could be used as trial functions to this kind of Hamiltonian. Therefore, it is not surprising that these coherent states let us to obtain easily a numerical description of the Dicke Hamiltonian. With these tool we intend to explore another observables of interest in the model.   

\section{Conclusion}
We have compared the numerical solutions for the ground and the first excited states for the finite Dicke Hamiltonian employing the Fock basis and the coherent basis. In the case of the Fock basis it was shown that both the computing time and the number of excitations in the field necessary for the convergence of the solution increases as the number of atoms do. In contrast, for the coherent basis the photon number needed to obtain convergence diminishes as the interaction parameter grows. Also, the expectation values of the principal observables agree with the ones in \cite{Ema03}. Finally, we have a tool which let us study more about the Hamiltonian, saving computing resources and tell us something about the system. 

JGH thanks B.M. Rodr\'\i guez-Lara for pointing out Ref. \cite{Che08}.  This work was partially supported by CONACyT-M\'exico, FONCICYT (project-94142), and DGAPA-UNAM .

\end{document}